\documentclass{article}
\usepackage{arxiv}
\usepackage[utf8]{inputenc} 
\usepackage[T1]{fontenc}    
\usepackage{hyperref}       
\usepackage{url}            
\usepackage{booktabs}       
\usepackage{amsfonts}       
\usepackage{nicefrac}       
\usepackage{microtype}      
\usepackage{lipsum}
\usepackage{natbib}
\usepackage{hyperref}
\hypersetup{colorlinks=true,
            linkcolor=blue,
            anchorcolor=blue,
            citecolor=blue}
\usepackage{graphicx}
\usepackage{amsmath}
\graphicspath{ {./images/} }
\usepackage{subcaption}
\usepackage{graphicx}
\usepackage{tabularx}

\usepackage{xcolor}

\title{Scaling Portfolio Diversification with Quantum Circuit Cutting Techniques}

\author{
 Vicente P. Soloviev \\ Fujitsu Research of Europe \\ \texttt{vicente.perezsoloviev@fujitsu.com} \\
 \And Antonio Márquez Romero  \\ Fujitsu Research of Europe \\ \texttt{antonio.marquezromero@fujitsu.com} \\
 \And Josh Kirsopp \\ Fujitsu Research of Europe \\ \texttt{josh.kirsopp@fujitsu.com} \\
 \And Michal Krompiec \\ Fujitsu Research of Europe \\ \texttt{michal.krompiec@fujitsu.com} \\
}

\begin{document}
\maketitle
\begin{abstract}
Quantum Approximate Optimization Algorithms (QAOA) have demonstrated a strong potential in addressing graph-based optimization problems. However, the execution of large-scale quantum circuits remains constrained by the limitations of current quantum hardware. In this work, we introduce \texttt{QuantCut}, an automatic framework for circuit cutting that enables efficient execution of large quantum circuits by decomposing entangling two-qubit gates into manageable sub-circuits. Specifically, we focus on gate-cutting techniques. We apply \texttt{QuantCut} to a 71-qubit QAOA circuit \textit{ansatz} for portfolio diversification in the S\&P 500 stock market, aiming to maximize asset diversification. Our approach iteratively optimizes the expectation value while leveraging circuit-cutting strategies to reduce the qubit register size. To validate our framework, we first conduct experiments on a toy model using quantum noise simulations for the Max-Cut problem, analyzing performance improvements with an increasing number of layers. Subsequently, we extend our methodology to a real-world financial optimization scenario, showing competitive results. The results suggest that \texttt{QuantCut} effectively facilitates large-scale quantum computations with circuit-cutting technologies.
\end{abstract}


\section{Introduction}
Quantum Approximate Optimization Algorithms (QAOA) \citep{farhi2014quantum} are widely used for graph-based optimization problems in the literature \citep{choi2019tutorial, blekos2024review}. 
QAOA involves optimizing a parameterized \emph{ansatz} to minimize a desired objective expectation value. In the interest of reducing the runtime demands, we use circuit-cutting techniques to decrease the width of the measurement circuits.

Circuit cutting \citep{simulating_large_quantum_circuits} consists in reproducing a quantum circuit by separating it into sub-circuits and reconstructing the final result with post-processing steps. Thus, each of the sub-circuits involves fewer qubits, allowing for larger computations on current quantum devices. Depending on how the original quantum circuit is partitioned, wire cutting and gate cutting have been proposed in the literature as distinct approaches. 

Gate cutting consists of four main steps: (i) identifying the optimal gate cuts to perform in the circuit; (ii) generating a subset of circuit experiments whose qubit counts are reduced compared to the full circuit; (iii) executing the experiments; and (iv) reconstructing the final result.

In this paper, we propose \texttt{QuantCut}: an automatic framework for circuit cutting which automatically applies the aforementioned steps to a given quantum circuit. We demonstrate it with a proof-of-concept application to maximization of the portfolio diversification among a set of assets selected from the S\&P 500 stocks \footnote{\url{https://www.kaggle.com/datasets/camnugent/sandp500}} with the QAOA approach by optimizing the expectation value of a 71-qubits \textit{ansatz}.

This paper is structured as follows: Section~\ref{sec_background} introduces some of the concepts and literature about QAOA, circuit cutting and portfolio diversification; Section~\ref{sec_method} introduces \texttt{QuantCut}, how the QAOA problem is decomposed and the data-engineering process for the data taken from the S\&P 500 stock market; Section~\ref{sec_results} presents the main results of the experimentation with real data and quantum noise; and Section~\ref{sec_conclussions} rounds the paper off with some further conclusions and future research lines.

\section{Background} \label{sec_background}

\subsection{Quantum approximate optimization algorithm} \label{backgroun_qaoa}
Variational quantum algorithms are hybrid classical-quantum approaches widely known in the literature~\citep{cerezo2021variational}. They involve three main ingredients: (i) an objective function to be minimized, (ii) a parametric quantum circuit (henceforth referred to as an \textit{ansatz}) and (iii) a classical optimizer that iteratively manipulates the \textit{ansatz}. Some examples are the Variational Quantum Eigensolver~\citep{peruzzo2014variational,tilly2022variational}, and the Quantum Approximate Optimization Algorithm (QAOA).

QAOA~\citep{farhi2014quantum} was initially proposed for combinatorial optimization problems. It has since been widely applied to challenges that can be represented as graphs, using quantum computing to explore large solution spaces. 

The QAOA \textit{ansatz} is composed of $n$ qubits and $p \in \mathbb{N}$ layers, internally composed by (i) a cost operator $U(H_C, \gamma)$ (Eq.~\ref{eq_U_c}) parameterized by $\gamma \in [0, 2\pi]$ encoding the optimization task to be solved ($H_C$), and (ii) the mixed operator $U(H_B, \beta)$ (Eq.~\ref{eq_U_b}) parameterized by $\beta \in [0, 2\pi]$, which represents a rotation in the $X$-axis of each qubit ($\sigma_j^x$).

\begin{equation}
    \label{eq_U_c}
    U(H_C, \gamma) = e^{-i \gamma H_C} = \prod^m_{\alpha=1}e^{-i\gamma C_{\alpha}},
\end{equation}
where $C_\alpha$ is the cost function to be minimized, $m$ the number of clauses that define the classical cost function to be optimized, and $H_C$ represents the cost Hamiltonian.

\begin{equation}
    \label{eq_U_b}
    U(H_B, \beta) = e^{-i \beta H_B} = \prod^n_{j=1}e^{-i\beta \sigma_j^x},
\end{equation}
where $H_B$ represents the mixed Hamiltonian.

\begin{figure}[]
\centering
\includegraphics[width=0.65\linewidth]{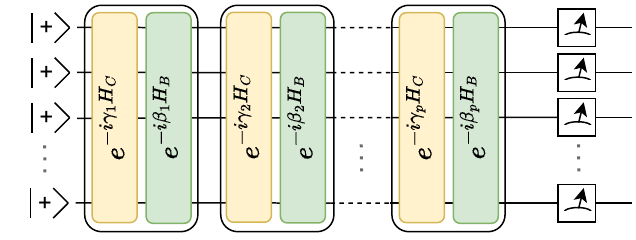}
\caption[QAOA \textit{ansatz} example]{\textit{ansatz} example with $p$ layers and $2p$ parameters. The quantum circuit is initialized from a superposition state over all possible computational basis states, then $p$ layers are applied, and all qubits are measured in the $Z$ basis.}
\label{qaoa_circ}
\end{figure}

Figure~\ref{qaoa_circ} shows an \textit{ansatz} example in which the qubits are initialized in a superposition of computational basis states, and iteratively applies $p$ layers ($2p$ parameters).

\subsection{Circuit cutting}
\label{background_circuit_cutting}
\begin{figure}[]
\centering
\includegraphics[width=0.4\linewidth]{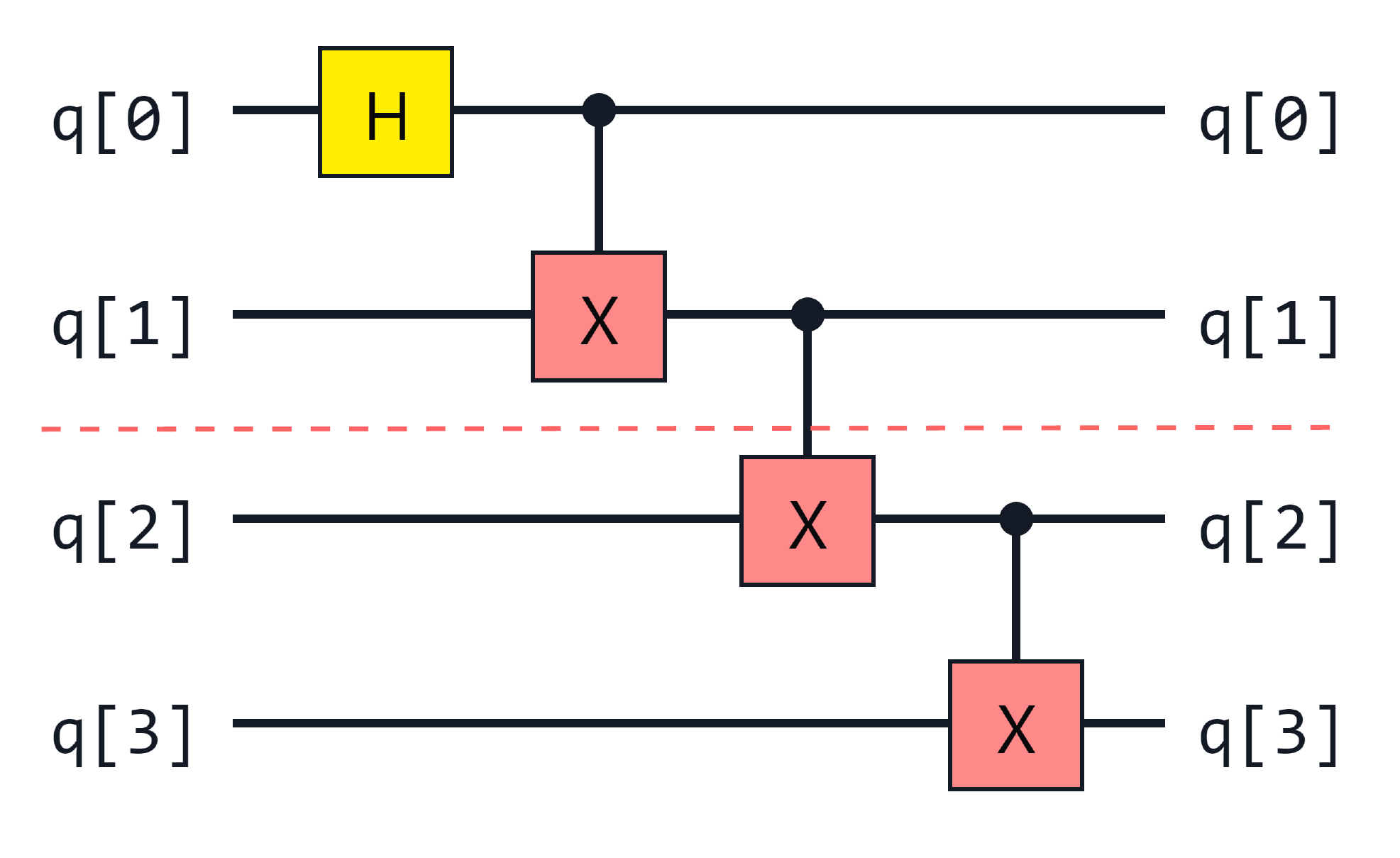}
\includegraphics[width=0.4\textwidth]{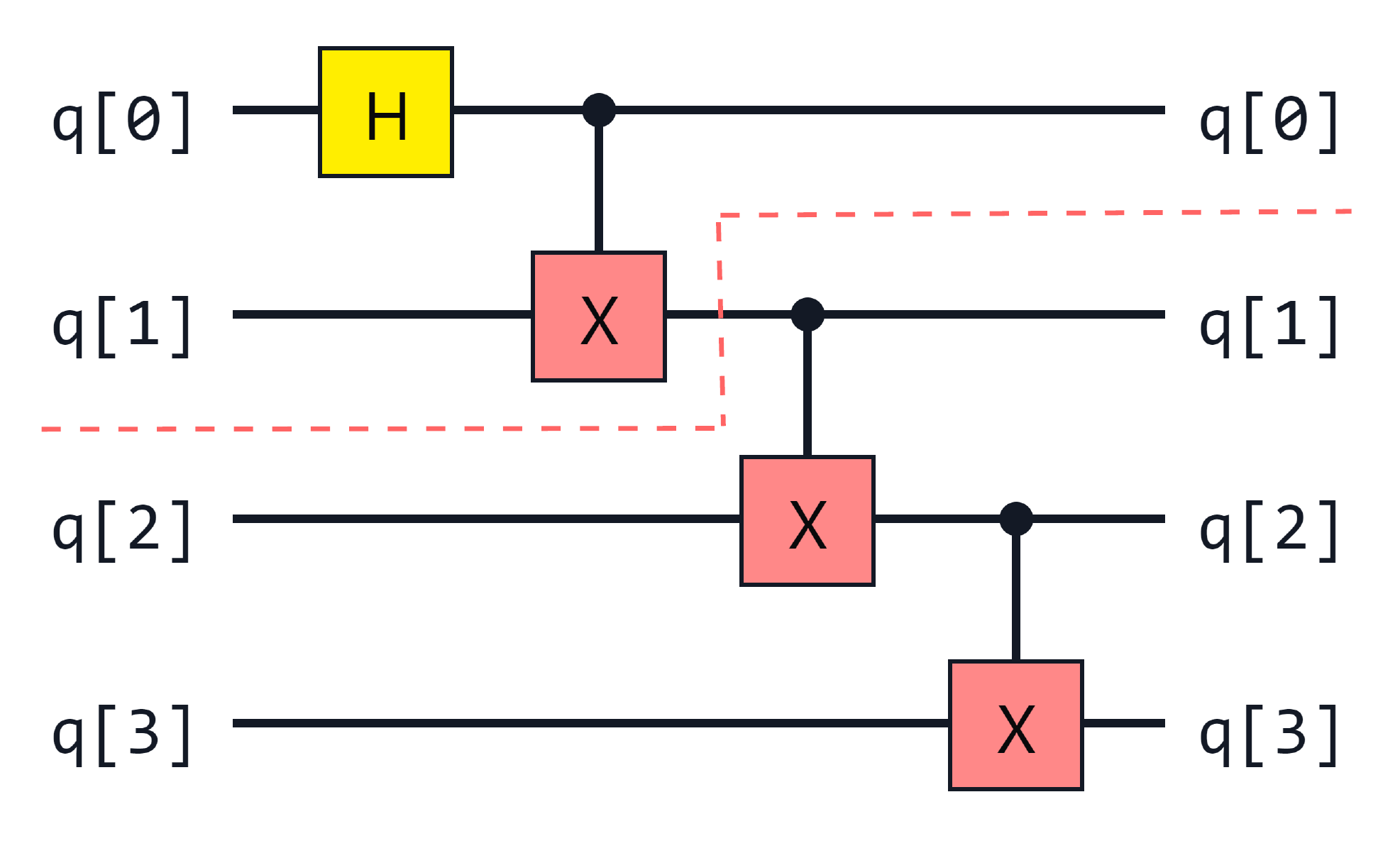}
\caption[Showcase visualization]{On the left, a gate-cutting (horizontal) cut applied to a GHZ circuit with four qubits. We observe that the second entangling CNOT is affected by the cut and its action will be reproduced in subsequent experiments in two subcircuits. On the right, a wire-cutting (vertical) cut is applied after the first entangling CNOT, the output quantum state will be saved and used as the initial state in the next circuit segment.}
\label{cuts}
\end{figure}

Circuit cutting is a technique designed to divide large quantum circuits into smaller, more manageable subcircuits, at the cost of performing more quantum experiments and additional classical post-processing overhead~\citep{simulating_large_quantum_circuits}. Such partitioning allows the quantum circuit to be run in different quantum devices in parallel, enabling distributed quantum computing~\citep{distributed,barral2025review,tejedor2025distributed}. Two flavors of cutting may be performed on a quantum circuit~\citep{optimal_partitioning}: wire- (also known as time-like, or vertical) and gate-cutting (also known as space-like, or horizontal). Their action on a quantum circuit can be seen in the diagram of Figure~\ref{cuts}.

Wire cutting involves partitioning a quantum circuit by cutting through the wires, or qubit lines, that evolve the quantum information over time~\citep{brenner2023optimal}. This method partitions the circuit by breaking the temporal continuity of the state, effectively creating segments that can be processed separately, for different time steps. 

Gate cutting, on the other hand, involves cutting through the entangling gates. This method partitions the circuit by breaking the qubit connections based on entanglement, resulting in smaller circuits where several experiments need to be run in order to reproduce locally the behavior of the entangling gates.

Both flavors incur an overhead scaling exponentially in terms of the number of cuts. In wire cutting, the overhead comes from the state preparation needed for the next circuit segment to be run, which is the measurement sampling result of the former segment. In gate cutting, the overhead comes from reproducing locally the non-local behavior of the entangling gates. In this work, we focus on the latter approach, and specifically for entangling gates involving two qubits only. 

Any two-qubit entangling gate $U$ can be represented as 
\begin{equation}\label{eq:KAK}
  U = [A_1 \otimes A_2] e^{i ( a_X X\otimes X + a_Y Y\otimes Y + a_Z Z\otimes Z) } [B_1 \otimes B_2],  
\end{equation}
where $X,Y,Z$ are the usual Pauli matrices and $A_1, A_2, B_1, B_2$ are one-qubit operators applied to numbered qubits $1$ and $2$, respectively. The interaction coefficients $[a_X, a_Y, a_Z]$ completely characterize the entangling gate. This representation is known as the KAK decomposition~\citep{tucci2005introduction}. We observe that, effectively, under this representation, cutting an entangling gate is equivalent to decomposing the \emph{interaction} term given in the two-qubit exponential. In \texttt{QuantCut}, we follow the procedure proposed by Mitarai and Fujii to automatically cut an interaction operator, involving generating six different circuits and post-processing the measurement results in Eq.~(B1) of~\citep{mitarai_fujii}. The channel $\mathcal{U}$ of the entangling unitary is reproduced by six different channels $\mathcal{F}$ involving only single-qubit gates
\begin{equation}\label{eq:qpd_channel}
    \mathcal{U} = \sum_{i=1}^6 a_i \mathcal{F}_i,
\end{equation}
where $a_i$ are coefficients related to the parameters in the interaction terms in Eq.~\ref{eq:KAK}. This technique is called \emph{quasiprobability decomposition}~\citep{qpdI,qpdII,qpdIII}, in the sense that we sample the probability distribution of a non-local entangling gate by means of a linear combination of the sampling results of local operations. 

We immediately observe that, if two or more entangling gates are to be cut, the post-measurement results need to be combined in order to reproduce the whole action of the entangling operators. Consequently, there is an exponential complexity as a function of the number of cuts involved in the circuit. 

The most common entangling gate found in quantum circuits is the CNOT gate, or controlled-$X$, whose error rate is also used as a common metric for the quality of the quantum computer. It can be written as 
\begin{equation}
\begin{split}
    CX &= e^{i \frac{\pi}{4} (I-Z) \otimes (I-X) }\\ &
    = e^{i \frac{\pi}{4} (II - IX - ZI + ZX) }\\&
    = e^{i \frac{\pi}{4}}  e^{-i \frac{\pi}{4} ( IX + ZI)} e^{i \frac{\pi}{4} ZX } \\& 
    = e^{-i \frac{\pi}{4}} [S \otimes (H S H)] e^{i \frac{\pi}{4} YI} e^{i \frac{\pi}{4} XX } e^{-i \frac{\pi}{4} YI}
\end{split}
\end{equation}
where we have omitted the Kronecker product $\otimes$ for notation simplicity, and where $H$ and $S$ are the standard Hadamard and phase gates, respectively. This decomposition involves one-qubit gates only, except for the $XX$ interaction term. Thus, we can follow the aforementioned recipe~\citep{mitarai_fujii} to cut the CNOT by cutting the interaction term $e^{i \frac{\pi}{4} XX }$. Further details may be found in Appendix~\ref{app:cnot}.

\subsection{Portfolio diversification}
\label{background_portfolio_diversification}
Portfolio diversification is a fundamental concept in financial management, aimed at reducing the risk inherently associated with investment, by allocating funds across a variety of assets. By combining assets with different risk profiles, expected returns, and correlations, investors can create a portfolio that balances potential gains with risk exposure.

Accordingly, we define the profit return of a given portfolio as,
\begin{equation}
    R_p = \sum^n_{i=1} w_iR_i,
\end{equation}
where $w_i$ and $R_i$ are the weight and expected return of asset $i$, respectively.

The portfolio variance, which represents its associated risk, is weighted through the pairwise covariance as,
\begin{equation}
\label{eq_variance_port}
    \sigma^2_p = \sum_{i=1}^n \sum_{j=1}^n w_i w_j \text{Cov}(R_i, R_j),
\end{equation}
where $\text{Cov}(R_i, R_j)$ is the covariance between the respective return of assets $i$ and $j$.

Portfolio diversification aims to mitigate the risk by minimizing the portfolio variance between the assets according to Equation~\ref{eq_variance_port}.

\section{\texttt{QuantCut}: An automatic tool for circuit cutting}
\label{quantcut}
In this section, we present our framework for automatic circuit cutting. More specifically, \texttt{QuantCut} implements an automatic gate-cutting strategy. The tool has been implemented in order to be easily integrated in the pipeline of other applications. Thus, it can automatically break the original quantum circuit into optimal sub-circuits conforming to the qubit number limitations of a given quantum device or simulator, and posteriorly reconstruct the measurement result of the full circuit. Moreover, it allows the user to fine tune each of the steps of the process. The main modules in the framework are: (i) the cut finder, which automatically returns the optimal gate cuts to be performed given an initial quantum circuit and respective maximum number of qubits allowed; (ii) the post processing, which generates the respective experiments according to the cuts (Section~\ref{background_circuit_cutting}) and posteriorly reconstructs the final result; and (iii) the plotting module for visualization of the results.

\begin{figure}[]
\centering
\includegraphics[width=0.8\linewidth]{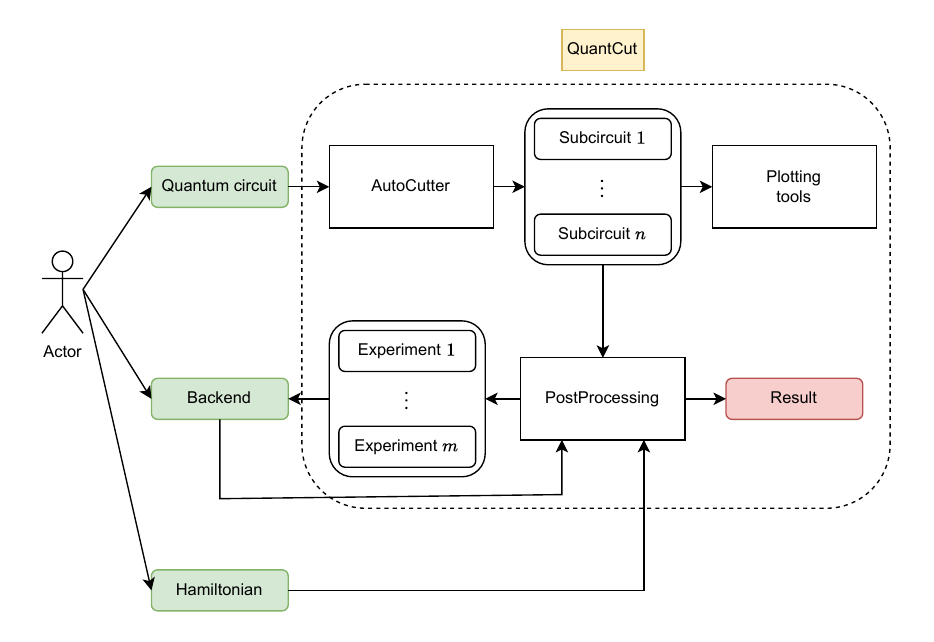}
\caption[QuantCut flowchart]{General flowchart of \texttt{QuantCut} where the three main modules are identified: (i) automatic cut finder (autocutter), (ii) post processing (postprocessing), and (iii) the plotting module (plotting tools).}
\label{quantcut_flowchart}
\end{figure}

Figure~\ref{quantcut_flowchart} shows a flowchart describing the behavior of \texttt{QuantCut}. The  inputs given by the user are (i) a quantum circuit to be cut, (ii) maximum number of qubits in subcircuits, (iii) a backend where the quantum circuits are executed (real quantum device or quantum simulator), and (iv) an (optional) operator to be measured on the quantum circuit. The following sections describe each of the cited modules in depth.

\subsection{Automatic cut finder}
\label{quantcut_cut_finder}
Given an original quantum circuit, the automatic cut finder will return the optimal cuts to be performed in order to minimize the number of cuts while satisfying the quantum device restrictions. 

Once the original quantum circuit is provided to the framework, manual and automatic cutting modes are available to the user. Each subcircuit generated by the framework acts upon a subset of qubits from the original quantum circuit. In order to perform circuit cutting, there must be at least two subcircuits with at least one qubit per subcircuit. The gates to be cut are those 2-qubit gates in which the control and target qubits are located in separate subcircuits. The more cuts to perform, the more complex the circuit-cutting procedure. Specifically, the number of experiments grows exponentially with the number of cuts. Thus, minimizing the number of cuts is critical. 

In this module, we perform the optimal placement of cuts and minimization of their number. The qubits' connectivities are represented as binary variables $x_{ij}$ where $x_{ij} = 0$ if there are no qubit gates between qubits $i$ and $j$, and $x_{ij} = 1$ otherwise. Hence, the optimization problem is formulated as:
\begin{align}
    & \text{minimize}\; f(x) \quad \text{subject to:} \nonumber \\
    & f(x) = 
    \begin{cases} 
        \sum_{i,j}^{i>j} x_{ij} w_{ij} & \text{if \texttt{isvalid}($x$)} \\ 
        \delta & \text{otherwise} 
    \end{cases},
\end{align}
where $x_{ij} \in {0, 1}$; $w_{ij}$ denotes the number of 2-qubit gates between qubits $i$ and $j$; \texttt{isvalid}($x$) is \emph{True} if the solution generates the subcircuits satisfying the maximum number of qubits constraint provided by the user, and \emph{False} otherwise; and $\delta$ is a penalization term. If a manual cut is requested instead of the automatic cut finder, then the positions of the gate cuts should be provided by the user.

\texttt{QuantCut} internally uses evolutionary computation to minimize the cost function. EDAspy~\citep{soloviev2024edaspy} is used to implement a version of the estimation of distribution algorithm here. This module returns the subcircuits as a result of the optimization process. To the best of our knowledge, this is the first disclosed implementation of an automatic algorithm that finds the most suitable gate cuts for gate-cutting strategies based on qubit connectivity.

\subsection{Post processing}
\label{quantcut_postprocessing}
This module generates a set of experiments from the given subcircuits and cuts (Section~\ref{quantcut_cut_finder}). Combining the results of these experiments, together with the given Hamiltonian or, generally, quantum operator to be measured, allows to reconstruct the final result. Note that the experiments are executed in a user-specified backend (quantum device or simulator)  where the original circuit can not be executed due to the number of qubits available. 

The Hamiltonian to be measured is split into subterms following the cuts given in the previous step. Then, the experiments are generated according to Section~\ref{background_circuit_cutting}. The expectation value for each of the experiments is independently computed. \texttt{QuantCut} parallelizes this computation over all the available CPUs (or QPUs, if available) to improve the runtime.

The framework provides not only the capability to return the reconstructed expectation value, but also the resulting state vector. Note that reconstructing the state vector is limited to a small number of qubits due to the exponential memory cost of storing the full state vector. 

\subsection{Plotting tool}

\begin{figure}[]
\centering
\includegraphics[width=0.70\linewidth]{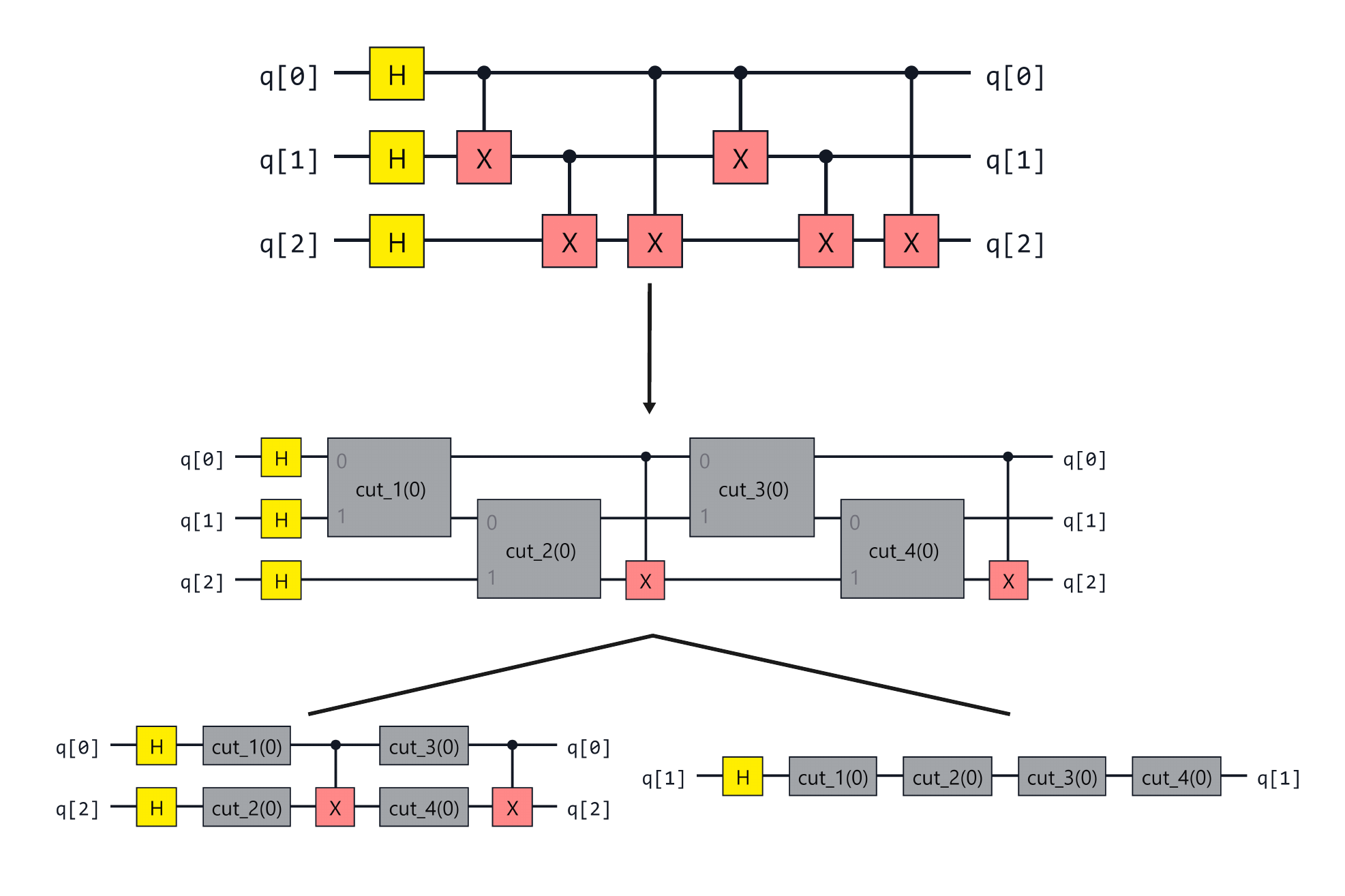}
\caption[Showcase visualization]{Example case in which an original quantum circuit is split into subcircuits of at most 2 qubits size.}
\label{quantcut_visualizaton_showcase}
\end{figure}

As previously mentioned, our framework provides graphical visualization of circuit cuts. Using \emph{pytket}’s rendering functionality~\citep{sivarajah2020t}, we depict the gates to be cut as gray boxes in the original quantum circuit. A custom \emph{pytket} gate is defined for this purpose, where each cut is labeled as \texttt{cut\_i} within a gray box, while the control and target qubits retain their configuration from the original gate. When visualizing the subcircuits separately, the gate cuts are displayed as single-qubit operations within each subcircuit. The labels are duplicated and positioned identically in the quantum circuit representation of each subcircuit. Figure~\ref{quantcut_visualizaton_showcase} shows an example in which an original quantum circuit is split into subcircuits of at most two qubits' size. Thus, the first subcircuit includes qubits “q[0]” and “q[2]”, while the second subcircuit includes qubit “q[1]”, respectively. 


\section{Method} \label{sec_method}
In this Section we explain how \texttt{QuantCut} (Section~\ref{quantcut}) has been combined with the QAOA approach (Section~\ref{backgroun_qaoa}) to target the portfolio diversification problem (Section~\ref{background_portfolio_diversification}).

The proposed approach follows the following pipeline:
\begin{enumerate}
    \item Data engineering (see Appendix~\ref{sec_method_data_eng} for further details). The initial data is preprocessed.
    \item Graph encoding (Section~\ref{sec_method_graph_encoding}). The stock market is represented as a graph in which nodes and edges represent the assets and covariances between them, respectively.
    \item Optimization task (Section~\ref{sec_opt_problem}). The QAOA circuit is constructed to solve the optimization problem. 
    \item Circuit cutting (Section~\ref{sec_circuit_cutting}). \texttt{QuantCut} is used to measure the original quantum circuit by performing circuit cutting.
    \item Solution extraction (Section~\ref{sec_solution_extraction}). After the optimal parameters are found, the optimal solution is extracted and analyzed.
\end{enumerate}

\subsection{Graph encoding} \label{sec_method_graph_encoding}
As described in Section~\ref{backgroun_qaoa}, QAOA has been traditionally applied to solve graph optimization problems. Thus, in this subsection we describe how to graphically encode the portfolio optimization problem.

The stock market and its graphical representation has been deeply studied in the literature \citep{nagurney2003innovations}. According to it, each node in the graph represents a stock asset in the market and the edges between the nodes represent the correlations in-between. In the interest of avoiding fully-connected graphs, the literature has proposed to impose a threshold $\alpha$, so that correlations below $\alpha$ will not be represented in the graph. This thresholding approach results in a market graph that more accurately reflects the actual structure of the stock market, where not all assets are significantly correlated with each other.\citep{nagurney2003innovations}.

\begin{figure}[]
\centering
\includegraphics[width=\linewidth]{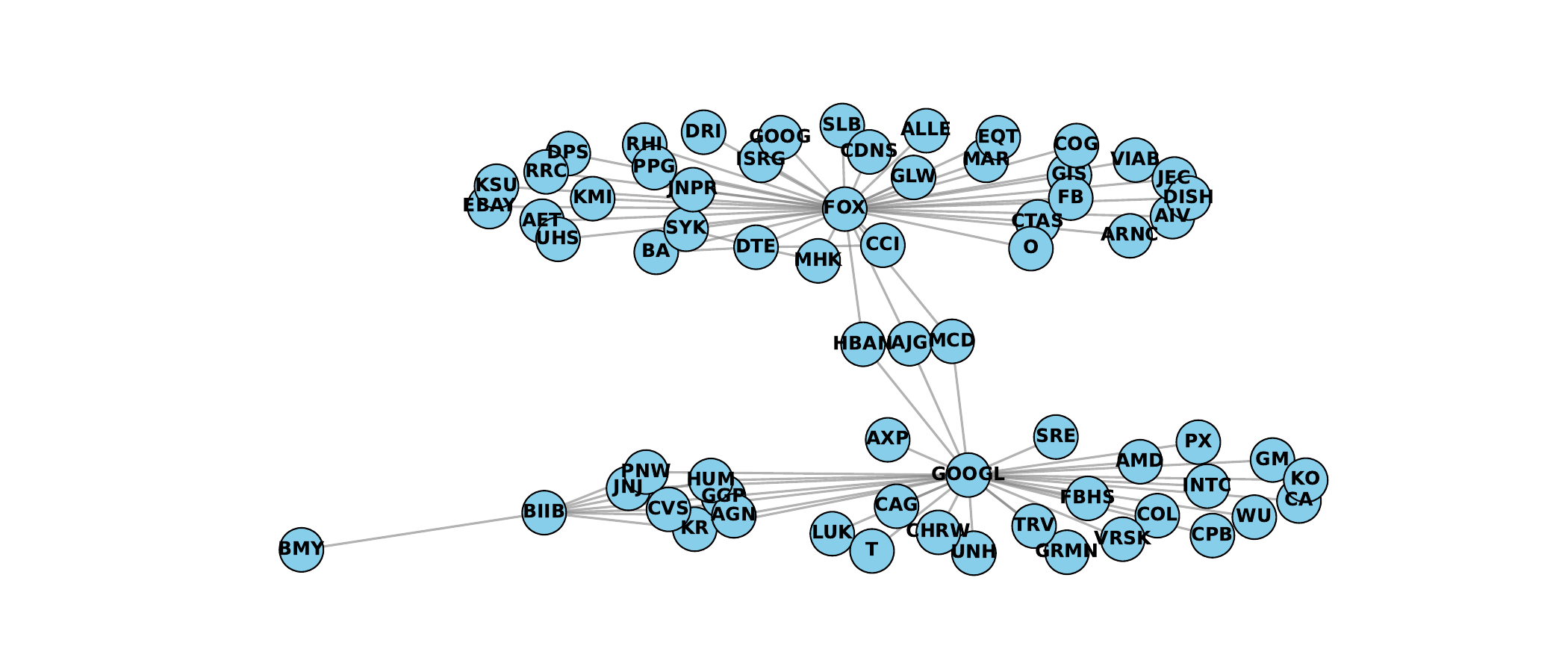}
\caption[Market graph]{Stock market graph where each node represents an asset in the market, labeled by their corresponding ticker symbols, and edges between them represent the pairwise correlations.}
\label{fig_stock_market_graph}
\end{figure}

In this work we will use one node per asset represented. Thus, we are using a graph composed of 71 nodes. The threshold $\alpha$ has been empirically adjusted to $\alpha=0.2$ over the covariance matrix (Section~\ref{sec_method_data_eng}). Figure~\ref{fig_stock_market_graph} shows the stock market graph used in our approach.

\subsection{Optimization problem} \label{sec_opt_problem}
This section describes how the diversification problem (Section~\ref{background_portfolio_diversification}) represented as a graph as described in Section~\ref{sec_method_graph_encoding} is solved by the QAOA approach.

Since the objective of the diversification problem is to reduce redundancy between highly correlated assets, the optimization task consists of detecting those conflicting edges. To this end, we have mapped the task into the Max-Cut problem, where nodes connected by maximal weights have to be assigned to different portfolios. 

The Max-Cut problem is NP-hard, which means that finding an exact solution for large graphs is computationally intensive and often impossible within a reasonable runtime. Approximation methods, heuristics, and algorithms such as semi-definite programming or quantum approaches (e.g. QAOA) are often employed to tackle it effectively. The optimization problem is defined as
\begin{equation} \label{eq_max_cut}
    \text{Max-Cut}(\text{G}) = \sum_{v_i, v_j \in E} w_{ij}x_i(1-x_j),
\end{equation}
where $E$ is the set of edges in the graph $\rm{G}$ from nodes $v_i$ to $v_j$ and associated weight $w_{ij}$; and $x_i$ and $x_j$ are binary variables which assign the portfolio to nodes $v_i$ and $v_j$, respectively. Since this is a binary optimization problem, the resulting solution of this process is the diversified strategy to locate the assets into two different portfolios. For more than two portfolios, the repeated bisection strategy \citep{dees2020portfolio} is used, in which the optimization task is repeatedly run over each of the sub-portfolios. In the extreme case where we infinitely many times run this strategy the result would be to assign each of the assets to a different portfolio.

As introduced in Section~\ref{backgroun_qaoa}, the market graph is mapped into the cost operator (Eq.~\ref{eq_U_c}) and applied in each of the layers of the \textit{ansatz}. In this approach, each node in the graph is assigned to a qubit in the system. There exist well-known strategies in the literature to map more than one node per qubit such as Pauli encoding \citep{sciorilli2025towards, maciejewski2024multilevel}. However, this was out of the scope of this project and is proposed as a future research direction. Each of the edges in the graph  involves either a combination of two (i) CX gates, or a (ii) CRZ gate between both qubits representing the respective nodes. In the interest of reducing the complexity of the circuit experiments in \texttt{QuantCut} we use the latter option in this approach.

\subsection{Circuit cutting} \label{sec_circuit_cutting}
This section describes how \texttt{QuantCut} is integrated in our QAOA pipeline, although it can be applied to other algorithms' pipelines.

For each parameter configuration in the \textit{ansatz}, circuit cutting is applied over the entire quantum circuit in order to compute the expectation value. The automatic cut finder (Section~\ref{quantcut_cut_finder}) is only run in the first iteration, and the same gate cuts are applied in the successive iterations. The post processing step (Section~\ref{quantcut_postprocessing}) is run in each of the iterations.

\subsection{Solutions extraction} \label{sec_solution_extraction}
Once we have an optimized quantum circuit, we would like to know the optimal solution, that is, the largest-amplitude qubit basis state. We could proceed by a straightforward measurement on all the wires, but, since we handle circuits with more than 70 qubits in this work, this task becomes unmanageable for a naive classical state vector simulation. An approximate solution may be found based on a dynamic-programming implementation of the marginal probabilities resulting from the measurement of subsets of qubits~\citep{tang2022cutting}. In this work, we opt for the simulation of the circuits using a tensor-network approach. 

Tensor networks have proven to be practical in many scientific areas involving high-dimensional numerical simulations~\citep{orus2014practical,banuls2023tensor,garcia2024survey}, including large quantum circuits~\citep{vidal2003efficient,markov2008simulating}. Broadly speaking, tensor networks are a collection of interconnected multidimensional matrices configured to obtain a quantity of interest through tensor contraction~\citep{evenbly2022practical}. They are an efficient representation in terms of the minimal units describing the problem at hand, allowing to keep the number of floating-point operations involved to a minimum. They become especially useful when dealing with sparse or low-correlated systems. Approximations on tensor networks can be made, allowing an exponential complexity reduction at the expense of a well-controlled accuracy.

\begin{figure}[ht]
  \centering
  \begin{subfigure}[t]{0.3\textwidth}
    \centering
    \includegraphics[width=\textwidth]{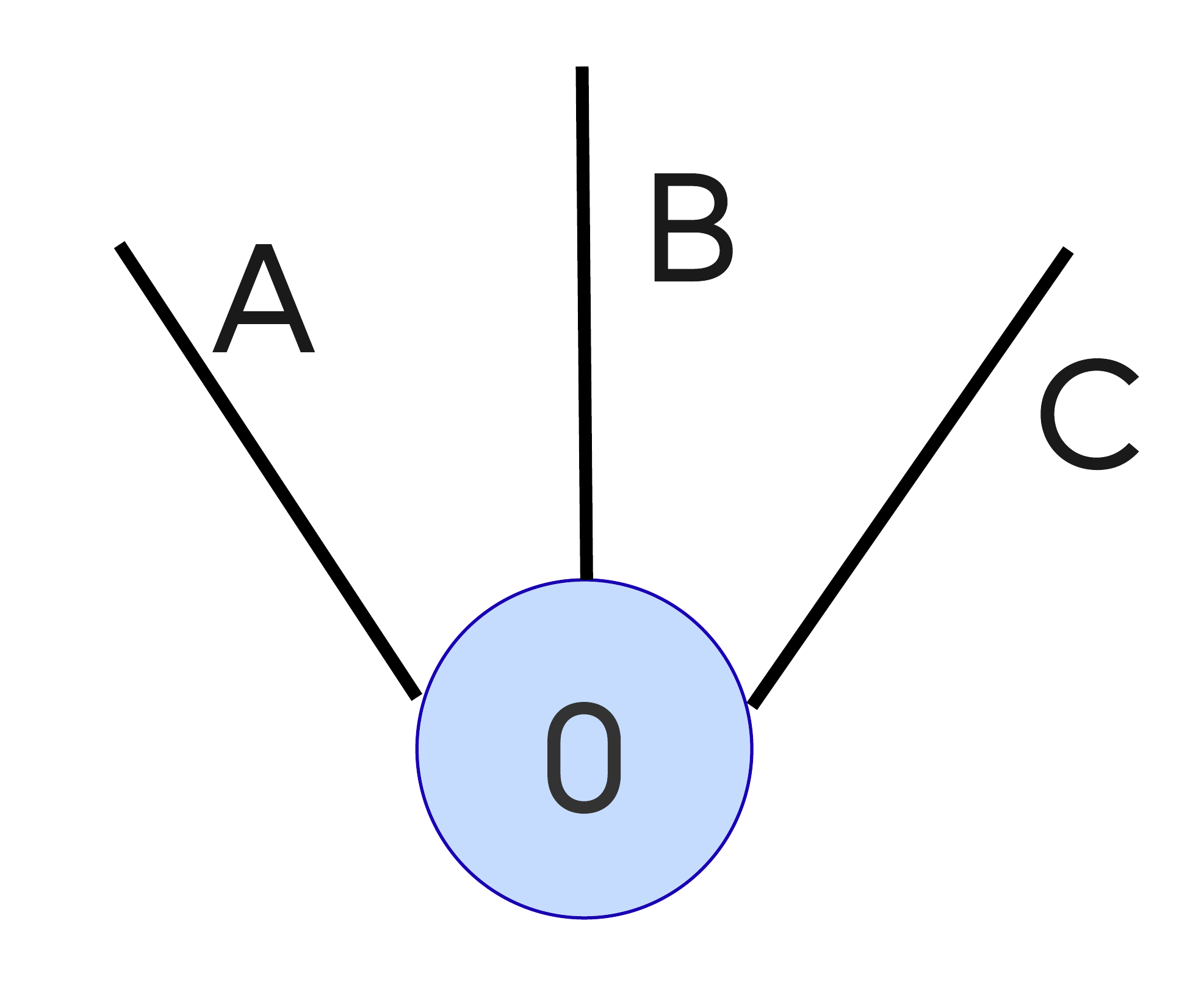}
    \caption{Depiction of a tensor with three indices, representing the quantum state $|000\rangle$.}
    \label{fig_0tensor}
  \end{subfigure}
  \hspace{0.05\textwidth}
  \begin{subfigure}[t]{0.4\textwidth}
    \centering
    \includegraphics[width=\textwidth]{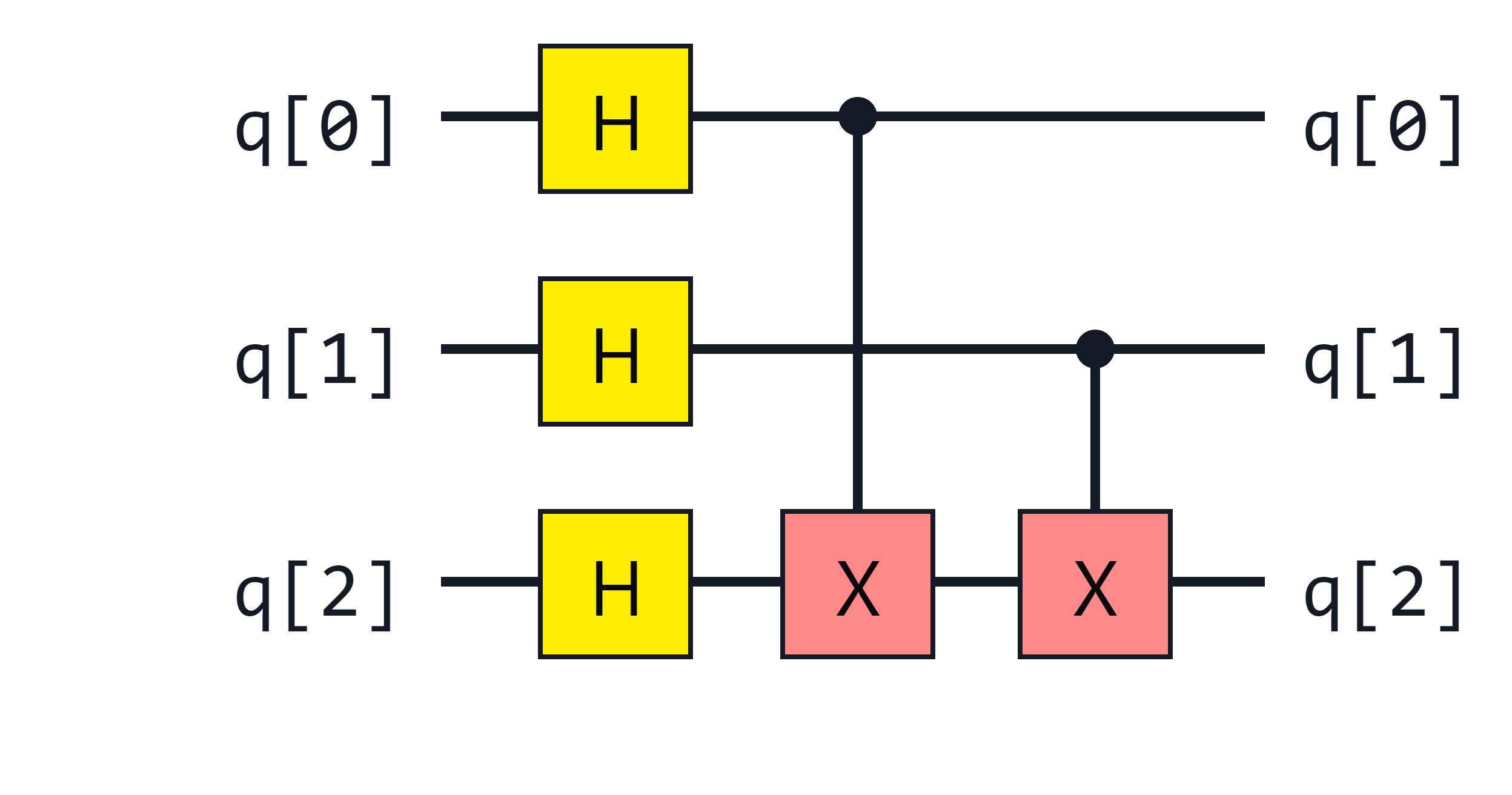}
    \caption{Instance of quantum circuit, consisting of three Hadamard gates and two CNOTs.}
    \label{fig_circuit}
  \end{subfigure}
  \vspace{5mm}
  \begin{subfigure}[t]{0.35\textwidth}
    \centering
    \includegraphics[width=\textwidth]{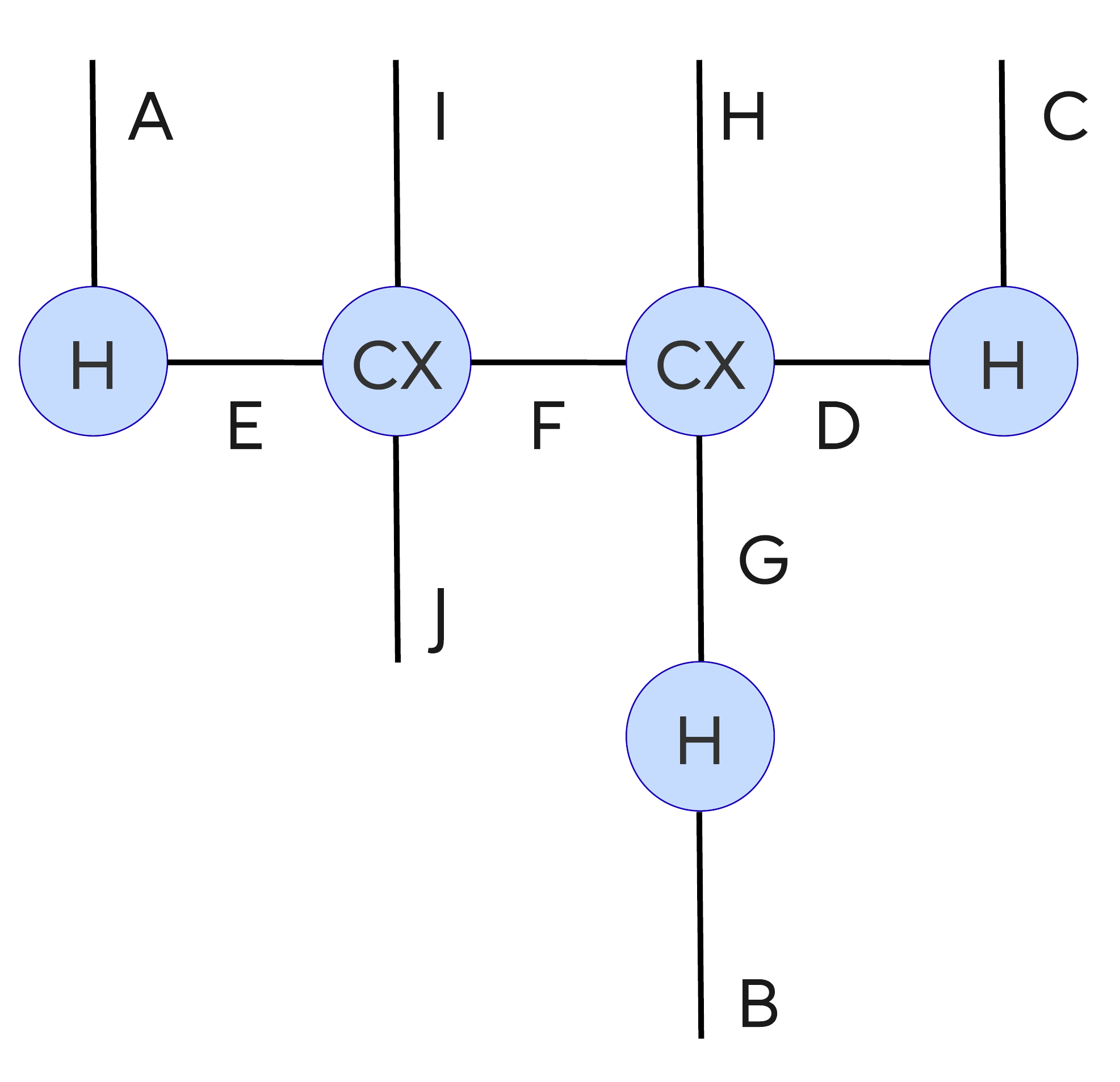}
    \caption{Tensor-network representation of the quantum circuit depicted in Fig.~\ref{fig_circuit}. Indices shared upon two gates are to be summed.}
    \label{fig_circuit_tensor}
  \end{subfigure}
  \hspace{0.05\textwidth}
  \begin{subfigure}[t]{0.4\textwidth}
    \centering
    \includegraphics[width=\textwidth]{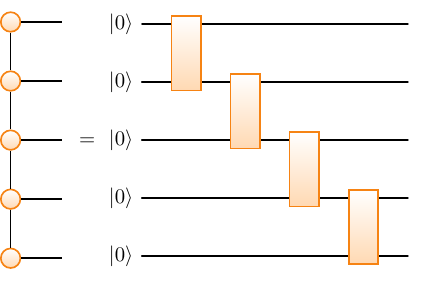}
    \caption{Matrix product state representation of a quantum circuit with neighboring entangling gates. Figure extracted from~\citep{garcia2024survey}.}
    \label{fig_mps_circuit}
  \end{subfigure}
  \caption{Tensor-network representation of quantum circuits and states.}
  \label{fig_tensors}
\end{figure}

We illustrate how a quantum circuit can be represented, and manipulated, by a tensor network. Let us start with the standard initial state in a quantum circuit, $|0\rangle$. For a circuit with three qubits, a standard representation for this state as a tensor is 
\begin{equation}\label{0state}
    |000\rangle \equiv \mathbf{0}_{ABC},
\end{equation}
where $A,B$ and $C$ are physical indices needed to extract a single element out of this quantum state. As we have three qubits and three indices, the dimensions of these are 2 for all three. Graphically, this tensor can be depicted as in Fig.~\ref{fig_0tensor}.
Now we apply the quantum circuit as described in Fig.~\ref{fig_circuit} to this state. Again, this circuit can be depicted as a tensor network as shown in Fig.~\ref{fig_circuit_tensor}. We observe six open indices: three for the input quantum state we have defined already in Eq.~\ref{0state} and another three for the resulting state. 

In terms of quantum operations, the resulting state is equivalent to 
\begin{equation}
\begin{split}
    |\Psi\rangle &= CX_{(1,2)} CX_{(0,2)}H_0H_1H_2 |000\rangle,\\ & 
    \equiv \mathbf{\Psi}_{HIJ},
    \end{split}
\end{equation}
and, in terms of tensor networks, the resulting quantum state is 
\begin{equation}
    \mathbf{\Psi}_{HIJ} = CX_{DFGH} CX_{EFIJ} H_{AE} H_{BG} H_{CD} \mathbf{0}_{ABC},
\end{equation}
where we follow Einstein's convention of summation: repeated indices are to be summed. We see that two-qubit gates involve tensors with four indices while one-qubit gates only require two. This expression is obtained by \emph{fusing} the tensor networks of the initial state in Fig.~\ref{fig_0tensor} and the quantum circuit in Fig.~\ref{fig_circuit_tensor}.

Thus, we observe how with the tensor-network formalism we are able to retrieve states from a quantum circuit. While for the aforementioned example we did not enforce it, there are specific tensor-network structures which allow a highly-optimized index contraction routine and an exponential reduction of computational resources needed to obtain such states under well-controlled approximations. One of the most popular representations is the \emph{matrix product state} (MPS)~\citep{perez2006matrix}, valid for quantum circuits involving only neighboring entangling gates as shown in Fig.~\ref{fig_mps_circuit}. If a quantum circuit has non-neighboring entangling gates, it needs to be transformed by inserting SWAP gates to conform to nearest-neighbour coupling.. While this introduces an additional overhead in the circuit depth, it is compensated by making use of the polynomially-scaling MPS representation of the quantum circuit. 


In \texttt{QuantCut}, we have used this tensor-network functionality as described above by means of the NVIDIA's \emph{cuQuantum}~\citep{bayraktar2023cuquantum} integrated in \emph{pytket}~\citep{sivarajah2020t} SDKs. In order to retrieve solutions, we sample the QAOA circuit and compute the associated amplitudes, extracting the solution with the largest amplitude. Calculations were made on a single NVIDIA A100 GPU.

\section{Results and discussion} \label{sec_results}
In this section, we firstly analyze the noise resilience of the approach with a toy example in which $n=10$ qubits are used (Section~\ref{sec_results_noise}), and secondly describe our results for portfolio diversification in which $n=71$ qubits are used (Section~\ref{sec_results_diversification}).

\subsection{Noise resilience} \label{sec_results_noise}
\begin{figure}[]
\centering
\includegraphics[width=\linewidth]{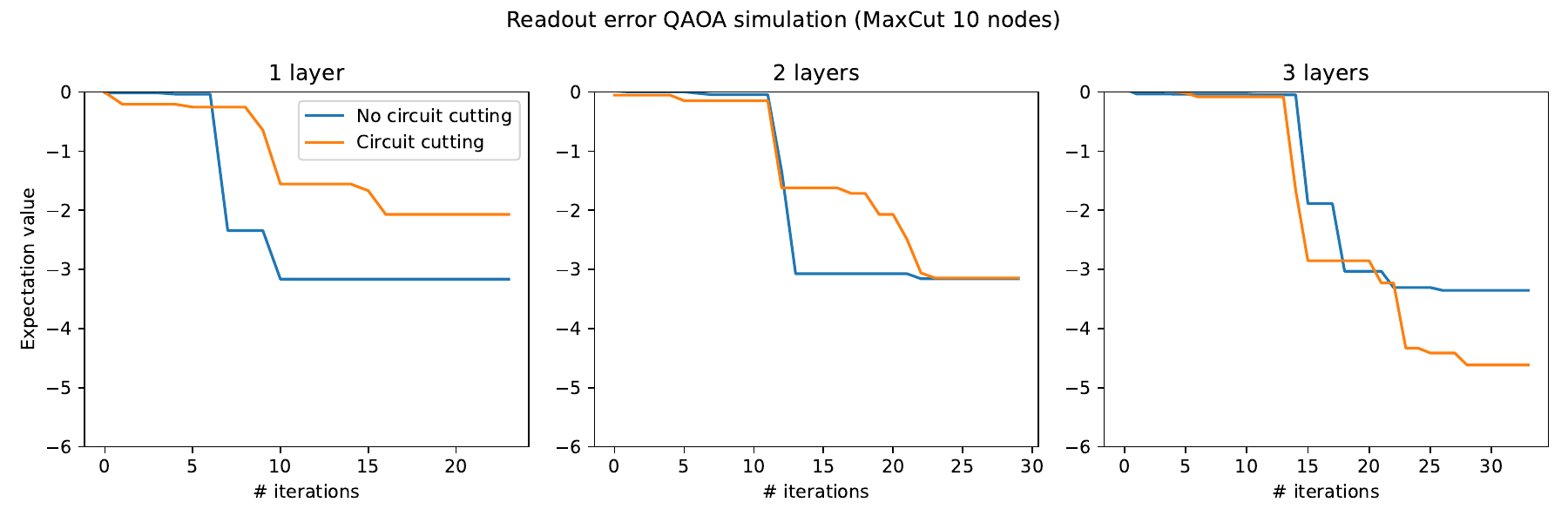}
\caption[Convergence plot noisy]{QAOA convergence plot with one, two and three layers, respectively, with (orange line) and without (blue line) circuit cutting.}
\label{fig_convergence_plot_noisy_toy}
\end{figure}

In this first experiment, we implement a toy example in which a random graph is generated and we solve the Max-Cut problem (Section~\ref{eq_max_cut}) over it by using the QAOA. We use Erdős-Rényi approach \citep{erdos1960evolution} to generate the initial graph with $n=10$ nodes. In this experiment, we will simulate a noisy environment by considering the readout noise model \citep{nielsen2010quantum} as,
\begin{equation}
    \begin{pmatrix}
    P(0|0) & P(0|1) \\
    P(1|0) & P(1|1)
    \end{pmatrix} = 
    \begin{pmatrix}
    0.99 & 0.01 \\
    0.01 & 0.99
    \end{pmatrix}.
\end{equation}
Figure~\ref{fig_convergence_plot_noisy_toy} shows a comparison of the expectation value convergence during runtime of the QAOA approach for $p\in\{1, 2, 3\}$ layers in a noisy environment. Runs with and without circuit cutting are compared in a noisy simulated environment. In this case, one gate cut is performed per layer in the \textit{ansatz}. A decreasing tendency of the converging expectation value is observed for an increasing number of layers in the QAOA \textit{ansatz}, as described in the literature: the more layers in the \textit{ansatz}, the better the expected performance. 
If the number of layers approaches infinity, the expectation value found is expected to be the optimal one.


Furthermore, a consistent improvement in circuit cutting accuracy is observed when compared to executions without cutting. However, with a single-layer configuration, the circuit cutting-based simulation demonstrates inferior convergence relative to its uncut counterpart. This trend reverses as the number of layers increases. These findings align with previous studies in the literature \citep{hakkaku2021sampling} and suggest promising directions for investigating circuit cutting as a viable error-mitigation strategy.

\subsection{71-qubit optimization} \label{sec_results_diversification}
In this second experiment, we solve the problem of portfolio diversification for 71-qubits, where \texttt{QuantCut} is used in each iteration to compute the expectation value.

\begin{figure}[]
\centering
\includegraphics[width=0.9\linewidth]{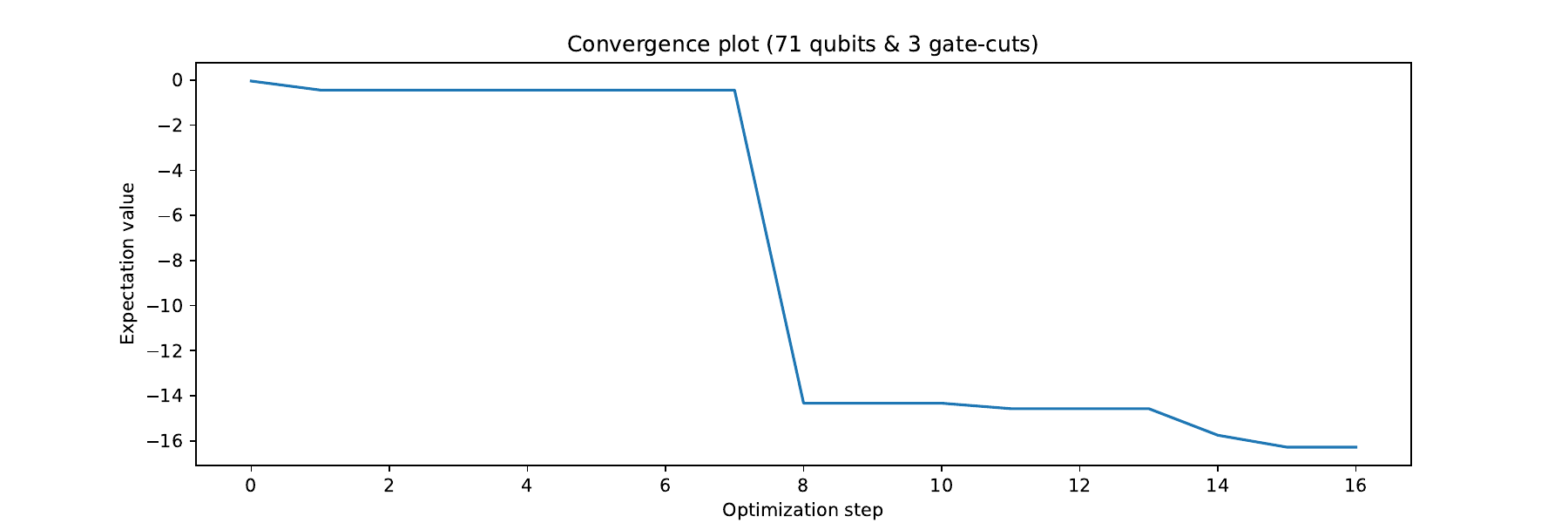}
\caption[Convergence plot]{Convergence plot: expectation value minimization of the loss function in Eq.~\ref{eq_max_cut} along the iterative optimization process.}
\label{fig_convergence_plot}
\end{figure}

Figure~\ref{fig_convergence_plot} shows the convergence plot in which \texttt{QuantCut} performs three gate cuts for layer in the QAOA \textit{ansatz}. Y- and X- axis represent the expectation value in each iteration and the runtime, respectively. A decreasing tendency is observed, showing that the classical optimizer is correctly minimizing the cost function. In this example, $p=1$ layers were used.
We now compare the results of our QAOA approach ($p=1$) with random sampling and with optimization performed with a classical optimizer. In the latter case, we will use an evolutionary algorithm (EA) \citep{soloviev2024edaspy}. Table~\ref{tab_metrics} shows a comparison of the cut edges found for the portfolio diversification problem (Eq.~\ref{eq_max_cut}) as well as the edges in between the found sub-portfolios which is desired to be minimum:
\begin{equation} \label{eq_acum_g_i}
    \text{Acum}(G_i) = \sum_{v_i, v_j \in G_i} w_{ij},
\end{equation}
where $G_i$ is a sub-portfolio generated as a result of the portfolio diversification problem over the original graph $G$, and $w_{ij}$ each of the edges between the nodes in the subgraph representation.

By maximizing the former (Eq.~\ref{eq_max_cut}) and minimizing the latter (Eq.~\ref{eq_acum_g_i}), the portfolio diversification is maximized. It is observed that our approach outperforms the random sampling. However, the EA approach achieved better results than QAOA. Based on the results found in Section~\ref{sec_results_noise} and literature precedents \citep{farhi2014quantum}, we suppose that by increasing the number of layers ($p$) our approach will potentially converge to similar solutions as the ones found by the EA.

\begin{table}[]
\centering
\caption{Random sampling, QAOA ($p=1$) and EA comparison of the Max-Cut($G$) (Eq.~\ref{eq_max_cut}) and $\text{Acum}(G_i)$ (Eq.~\ref{eq_acum_g_i}) metrics, where $G_i$ are the resultant subgraphs after performing the cuts.}
\vspace{1em}

\begin{tabularx}{0.8\textwidth}{|X|X|X|X|}
\hline
& Max-Cut($G$) & Acum($G_0$) & Acum($G_1$) \\
\hline
Random sampling & $26.1 \pm 2.1$ & $15.2 \pm 5.0$ & $7.9 \pm 7.1$ \\
QAOA ($p=1$) & 43.41 & 2.439 & 4.113 \\
EA & 46.51 & 1.190 & 2.110 \\
\hline
\end{tabularx}

\label{tab_metrics}
\end{table}

\section{Conclusions} \label{sec_conclussions}
In this work, we present \texttt{QuantCut}: an automatic tool for circuit cutting. The tool is able to automatically find optimal locations of gate cuts by minimizing the number of cuts, and to perform post-processing calculations to reconstruct the measurement of the original circuit. \texttt{QuantCut} can be applied for the reconstruction of expectation values of operators and state vectors, although for the latter we must address the exponential memory-storage cost. 

We iteratively use \texttt{QuantCut} to compute the expectation value of the QAOA \textit{ansatz} during the optimization runtime. Firstly, we apply this pipeline to a simple example of a simulation with a quantum noise model, for the Max-Cut optimization problem, and analyze its behavior with respect to increasing the number of layers. Secondly, we solve a proof-of-concept problem in the finance domain: we maximize the diversification among selected assets from the S\&P 500 stock market. In this work, we simulate up to 71-qubit quantum circuits. After optimization of the QAOA circuits, solutions in the form of a qubit-basis state can be obtained by sampling the circuits. Matrix product state tensor network representations of the circuits were used for optimal sampling performance. The numerical results demonstrate that our approach provides a competitive workflow for this problem, and may potentially increase resilience to quantum noise.

While this work has focused on QAOA and its applications to finance, we expect \texttt{QuantCut} to be a versatile tool with widespread applications where the simulation of large quantum circuits is involved. Specially interesting is the application of this automatic pipeline in the area of distributed quantum computing.

As future research work we propose to: (i) apply other problem mappings for assigning more than one asset per qubit; (ii) increase the number of layers in the QAOA \textit{ansatz}; and (iii) analyze the performance of the approach and resilience of circuit cutting when different types of quantum noise are present. 

\newpage
\appendix

\section{Cutting a CNOT}\label{app:cnot}

In this Appendix, we provide thorough details to perform the cut of the most common entangling gate found in quantum circuits: the controlled-$NOT$, or controlled-$X$, gate. It is well known that the CNOT, or $CX$ can be written as 
\begin{equation}
    CX = e^{i \frac{\pi}{4} (I-Z) \otimes (I-X) },
\end{equation}
where we can develop the terms to find out that
\begin{equation}
\begin{split}
    CX &= e^{i \frac{\pi}{4} (I-Z) \otimes (I-X) }\\ &
    = e^{i \frac{\pi}{4} (II - IX - ZI + ZX) }\\&
    = e^{i \frac{\pi}{4}}  e^{-i \frac{\pi}{4} ( IX + ZI)} e^{i \frac{\pi}{4} ZX } \\& 
    = e^{-i \frac{\pi}{4}} [S \otimes (H S H)] e^{i \frac{\pi}{4} YI} e^{i \frac{\pi}{4} XX } e^{-i \frac{\pi}{4} YI} \\&
    = \frac{e^{-i \frac{\pi}{4}}}{2\sqrt{2}} [S \otimes (H S H)] [(I + iY)\otimes I] e^{i \frac{\pi}{4} XX } [(I-iY)\otimes I] 
\end{split}
\end{equation}
where we have at times omitted the Kronecker product $\otimes$ for the brevity of notation, and $H$ and $S$ are the standard Hadamard and phase gates, respectively. We have developed the expression to the minimal units separable in products of single-qubit gates applied to the action qubits, apart from the non-separable interaction operator $e^{i \frac{\pi}{4} XX }$. 

Now, we can follow the aforementioned algorithm of reference~\citep{mitarai_fujii} to cut the CNOT by cutting the interaction operator, generating six different subexperiment circuits with only single-qubit gates. In addition, a practical guide into circuit cutting is also given on \emph{Qiskit}'s explanatory material~\citep{qiskit-addon-cutting}, where they introduce an example for cutting the $R_{ZZ}$ gate.

We summarize the aforementioned implementation using a general two-qubit unitary of the form 
\begin{equation}
    R = e^{i \theta A_1 \otimes A_2},
\end{equation}
with $A_i$ being any Pauli operator. We notice that the exponential sum in Eq.~\ref{eq:KAK} can always be decomposed in the product of exponentials because the set of Pauli words $XX,YY,ZZ$ commutes all with each other. The action of this gate on a two-qubit state generates sets of measurement counts which we denote by the general channel term $\mathcal{U}$. According to the algorithm, we have to generate six subexperiments to replicate this channel as Eq.~\ref{eq:qpd_channel} with only single-qubit gate channels $\mathcal{F}_i$. In this case, the coefficients $a_i$ are given by 
\begin{equation}
\begin{split}
    a_1 &= \cos(\theta)^2,\\
    a_2 &= \sin(\theta)^2,\\
    a_3 &= \sin(2\theta)/2,\\
    a_4 &= -a_3,\\
    a_5 &= a_3,\\
    a_6 &= -a_3.
\end{split}
\end{equation}

We list now the six subexperiments needed:
\begin{enumerate}
    \item We do nothing to the input circuit, we apply identity gates $I \otimes I$ and sum the contribution as 
    \begin{equation}
        \mathcal{U} \longrightarrow \mathcal{U} + a_1 \mathcal{F}_1.
    \end{equation}
    We notice that $\mathcal{F}_1$ is the measurement result after the action of the two identity gates, that is, the quantum circuit without the entangling gate applied.

    \item We apply the operators $A_1 \otimes A_2$, count the results in $\mathcal{F}_2$ and add the contribution 
    \begin{equation}
        \mathcal{U} \longrightarrow \mathcal{U} + a_2 \mathcal{F}_2.
    \end{equation}

    \item We apply the operators $M_{A_1} \otimes e^{i \frac{\pi}{4} A_2}$, where $M_{A_1}$ is a projective measurement onto the $A_1$ basis. Thus, an additional classical register is needed for this subexperiment. Also, since we can only measure in the $Z$ basis, if $A_1$ is $X$ or $Y$, we need to rotate to this basis first, that is, 
    \begin{equation}
    \begin{split}
        M_{X} &= H M_{Z},\\
        M_{Y} & = S^{\dagger} H M_{Z}. 
    \end{split}
    \end{equation}
    Associated with this measurement is the parameter $\beta$, with values $1$ or $-1$ if the projective result of the first qubit is $0$ or $1$, respectively.
    We recollect the measurement results from this experiment in $\mathcal{F}_3$ and update as 
    \begin{equation}
        \mathcal{U} \longrightarrow \mathcal{U} + \beta a_3 \mathcal{F}_3.
    \end{equation}  

    \item Similar to experiment 3, but now with the gates $M_{A_1} \otimes e^{-i \pi/4 A_2}$, and update 
    \begin{equation}
        \mathcal{U} \longrightarrow \mathcal{U} + \beta a_4 \mathcal{F}_4.
    \end{equation} 

    \item We apply the operators $e^{i \frac{\pi}{4} A_1} \otimes M_{A_2}$, and update 
    \begin{equation}
        \mathcal{U} \longrightarrow \mathcal{U} + \beta a_5 \mathcal{F}_5.
    \end{equation} 

    \item Similar to experiment 5, with gates $e^{-i \frac{\pi}{4} A_1} \otimes M_{A_2}$, and update 
    \begin{equation}
        \mathcal{U} \longrightarrow \mathcal{U} + \beta a_6 \mathcal{F}_6.
    \end{equation}     
        
\end{enumerate}

After these six experiments, count recollection in $\mathcal{U}$ will be approximately the same as the original recollection from the two-qubit unitary action on the input qubit state. For the specific case of the CNOT, $\theta = \frac{\pi}{4}$, so the quasiprobabilities become $[a_1, a_2, a_3, a_4, a_5, a_6] = [0.5, 0.5, 0.5, -0.5, 0.5, -0.5]$. $A_1=A_2=X$ and thus we need to perform $M_X$ measurements in steps 3-6. 

\newpage
\section{Data engineering}\label{sec_method_data_eng}

For this experiment we will be using real data from the S\&P 500 stock market~\footnote{\url{https://www.kaggle.com/datasets/camnugent/sandp500}}, which collects the stock values from the last five years. In the interest of showcasing a real scenario, 71 assets out of 500 were 
carefully selected. 

\begin{figure}[]
\centering
\includegraphics[width=\linewidth]{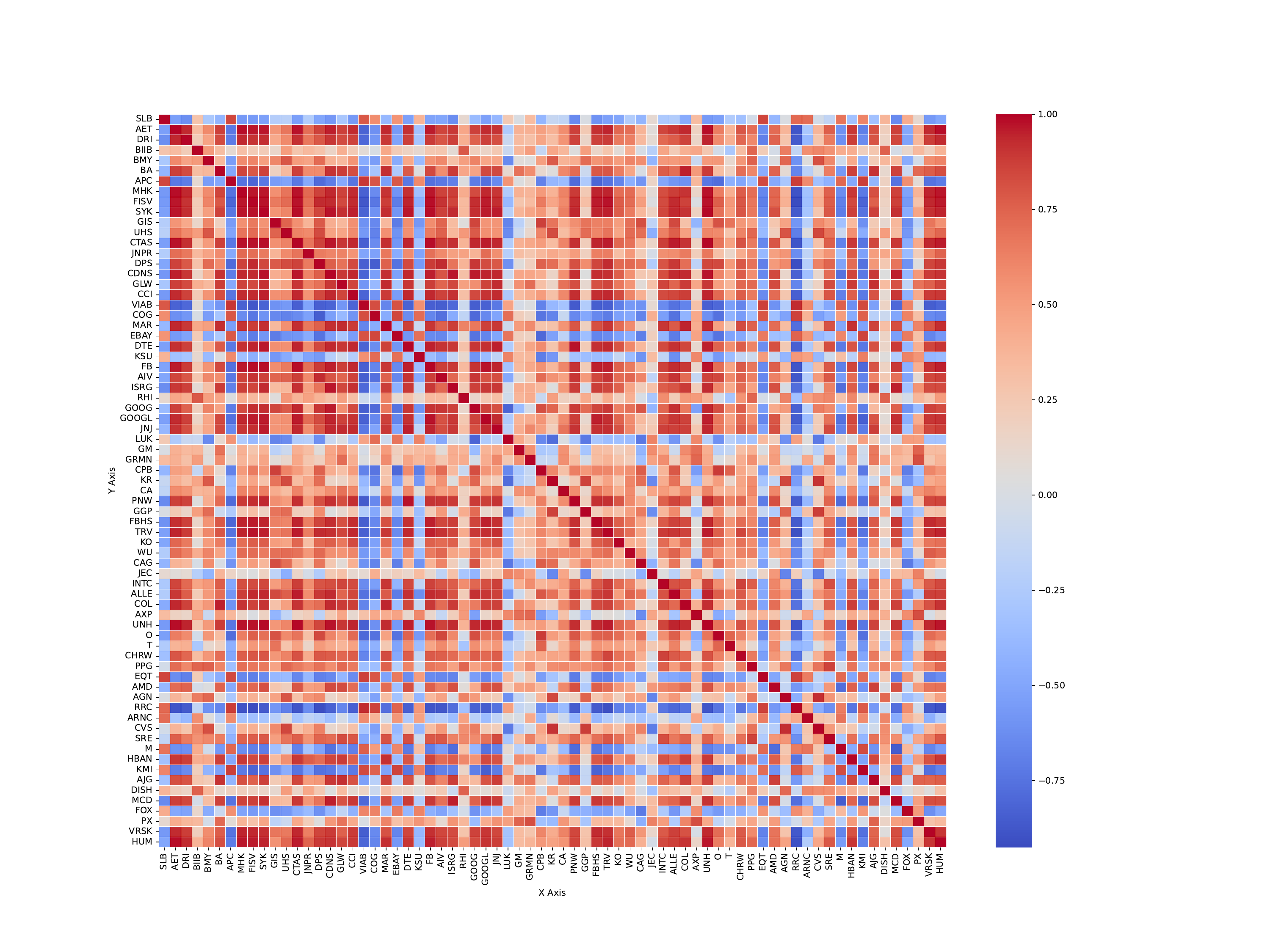}
\caption[Correlation heatmap]{Correlation heatmap for all 71 assets considered in this work, where positive and negative correlations are represented as light and dark colors, respectively.}
\label{fig_heatmap}
\end{figure}

In this dataset we first analyze the correlation between the time series representing the time evolutions of the assets. Figure~\ref{fig_heatmap} shows a correlation heatmap among all the selected assets in which light and dark colors represent the positive and negative Pearson correlation coefficients, respectively, as
\begin{equation}
    r = \frac{\sum (x_i y_i) - n \mu_x \mu_y}{\sqrt{\left( \sum x_i^2 - n \mu_x ^2 \right) \left( \sum y_i^2 - n \mu_y ^2 \right)}},
\end{equation}
where $x_i$ and $y_i$ represent the individual data points in two different datasets and $\mu_x$ and $\mu_y$ represent the mean values of both datasets. 

\begin{figure}
\centering
\begin{subfigure}{0.45\textwidth}
    \includegraphics[width=\textwidth]{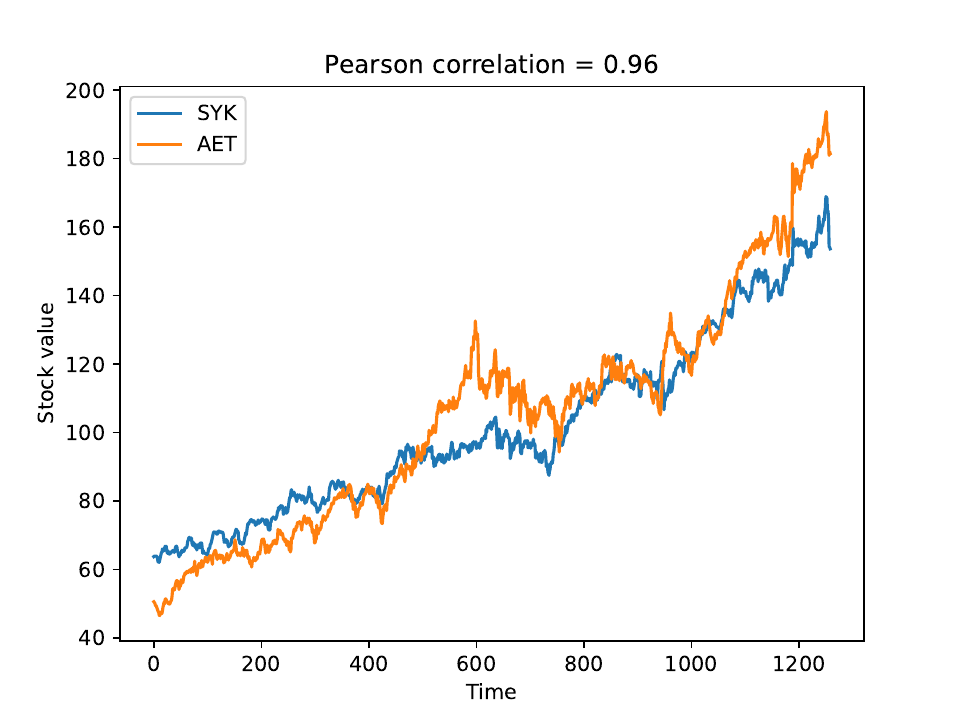}
    \caption{Time series representing two stock values along time with positive correlation.}
    \label{fig_time_series_postitive_corr}
\end{subfigure}
\hspace{0.05\textwidth}
\begin{subfigure}{0.45\textwidth}
    \includegraphics[width=\textwidth]{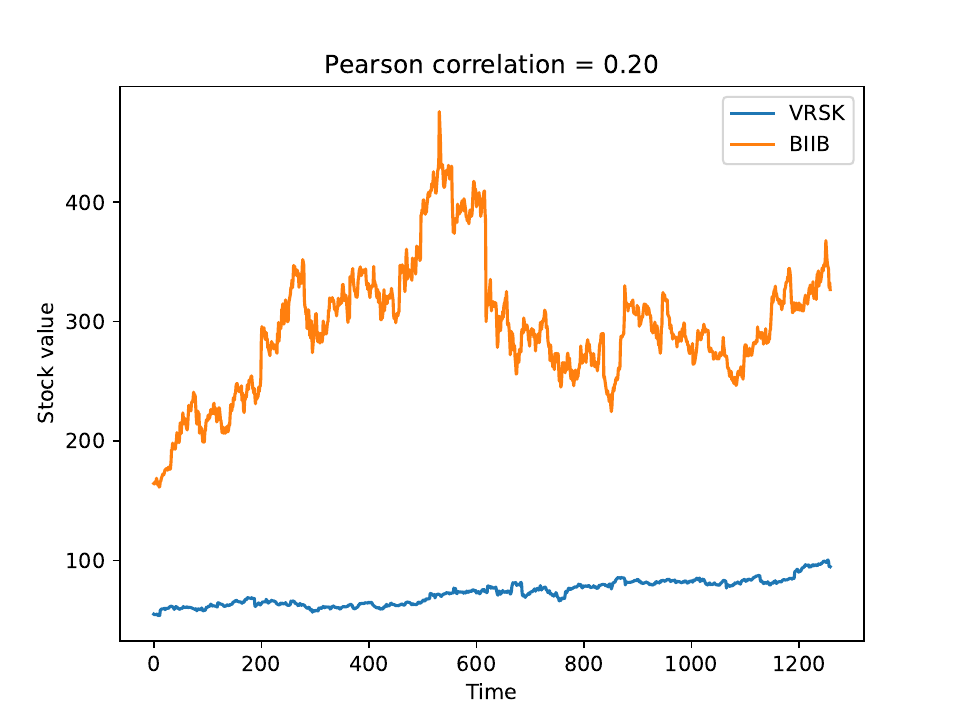}
    \caption{Time series representing two stock values along time with low correlation.}
    \label{fig_time_series_low_corr}
\end{subfigure}
\vspace{5mm}
\begin{subfigure}{0.45\textwidth}
    \includegraphics[width=\textwidth]{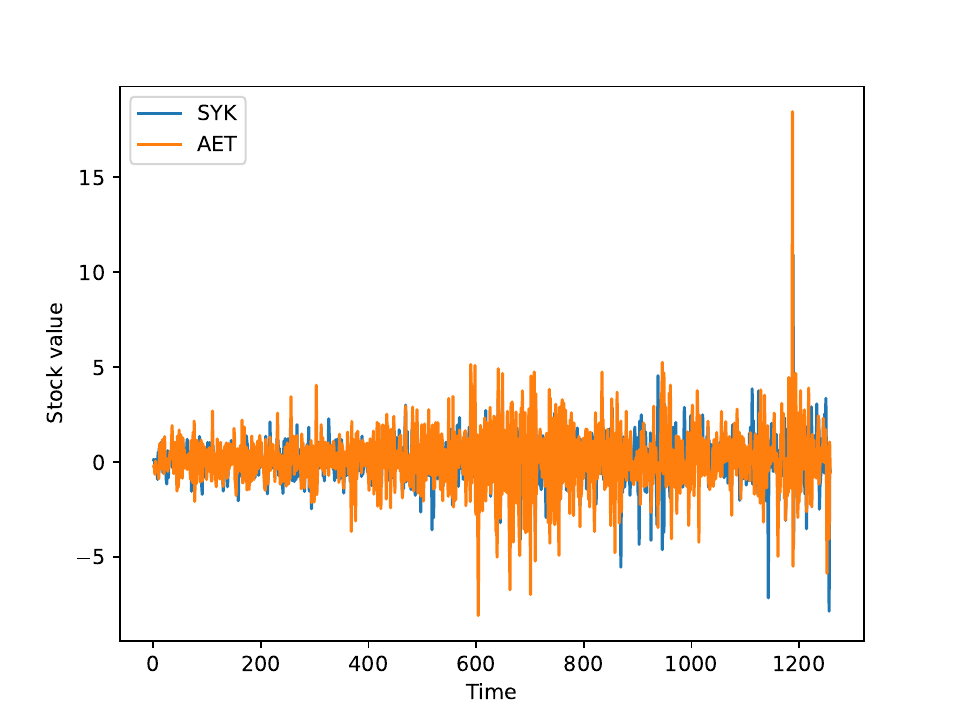}
    \caption{Rolling window time series representing two stock values along time with positive correlation.}
    \label{fig_time_series_postitive_corr_stand}
\end{subfigure}
\hspace{0.05\textwidth}
\begin{subfigure}{0.45\textwidth}
    \includegraphics[width=\textwidth]{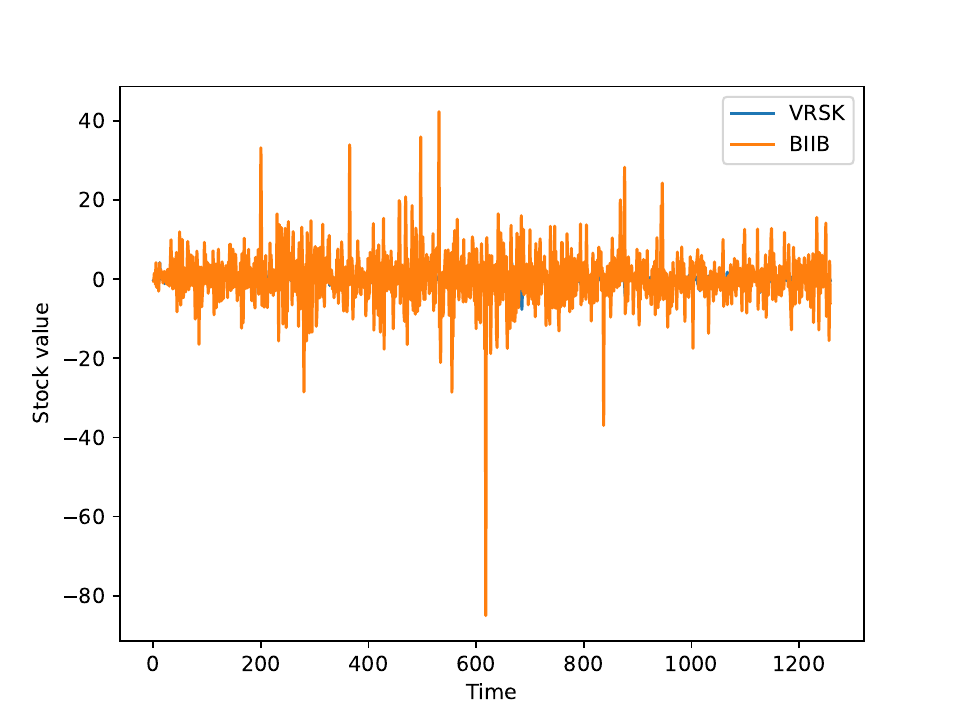}
    \caption{Rolling window time series representing two stock values along time with low correlation.}
    \label{fig_time_series_low_corr_stand}
\end{subfigure}
\caption{Time series examples plotting along time.}
\label{fig_time_series}
\end{figure}

A positive correlation between two temporal time series of stock assets occurs when the movements of the two assets tend to align in the same direction over time. This means that as the value of one stock increases (decreases), the value of the other stock is likely to increase (decrease) in a similar manner. Figure~\ref{fig_time_series} shows one example in which two assets follow a positive correlation along time (Subfigure~\ref{fig_time_series_postitive_corr}), and another example in which two assets are not correlated at all (Subfigure~\ref{fig_time_series_low_corr}).


In order to pre-process the time series, we normalize and standardize the data. Firstly, we normalize the data between 0 and 1 by scaling each each data point as $x' = \frac{x - x_{\rm{min}}}{x_{\rm{max}} - x_{\rm{min}}}$ where $x_{\rm{min}}$ and $x_{\rm{max}}$ are the minimum and maximum values detected in the time series, respectively. Secondly, standardization transforms the data to have a mean of $\mu'' = 0$ and a standard deviation of $\sigma'' = 1$ by computing each data point as $x'' = \frac{x' - \mu'}{\sigma'}$, where $\mu'$ and $\sigma'$ are the mean and standard deviation of the time series after the running the normalization, respectively.


After pre-processing the data series, we compute the covariance matrix over the standardized data, in which off-diagonal elements represent the pairwise variances among assets.

\newpage
\bibliographystyle{plainnat}
\bibliography{references}

\end{document}